\documentclass[fleqn,usenatbib,useAMS]{mnras}
\usepackage{graphicx}   
\usepackage{amsmath}    
\usepackage{amssymb}    
\usepackage{multicol}        
\usepackage{bm}     
\usepackage{pdflscape}  
\usepackage{longtable}

\usepackage[T1]{fontenc}
\usepackage{ae,aecompl}

\title[Statistic of Supra-Arcade Downflows]{Statistical Investigation of the Kinematic and Thermal Properties of Supra-arcade Downflows Observed During a Solar Flare}

\author[Tan et al.]{
Guangyu Tan$^{1}$,
Yijun Hou$^{2,3}$\thanks{Corresponding author, \href{mailto:yijunhou@nao.cas.cn}{yijunhou@nao.cas.cn}},
and Hui Tian$^{1,2}$\thanks{Corresponding author, \href{mailto:huitian@pku.edu.cn}{huitian@pku.edu.cn}}
\\
$^{1}$School of Earth and Space Sciences, Peking University, Beijing 100871, China
\\
$^{2}$National Astronomical Observatories, Chinese Academy of Sciences, Beijing, 100101, China
\\
$^{3}$School of Astronomy and Space Science, University of Chinese Academy of Sciences, Beijing, 100049, China
}
\pubyear{2022}
\begin{document}
\label{firstpage}
\pagerange{\pageref{firstpage}--\pageref{lastpage}}
\maketitle

\begin{abstract}
Supra-arcade downflows (SADs) are dark structures descending towards post-reconnection flare loops observed in extreme ultraviolet or X-ray observations and are closely related to magnetic reconnection during solar flares. Due to the lack of statistical study on SADs in a single flare, evolutions of kinematic and thermal properties of SADs during the flare process still remain obscure. In this work, we identified 81 SADs in a flare that occurred on 2013 May 22 using observations of the Atmospheric Imaging Assembly (AIA) on the Solar Dynamics Observatory (SDO). The kinematic properties of each SAD, including the appearance time, height, projective velocity, and acceleration were recorded. We found that the appearance heights of SADs become larger during the flare, which is likely due to the lift of the bottom of the plasma sheet. In the flare decay phase, the region where SADs mainly appear moves from the north part to the south side possibly related to a secondary eruption in the south side. The trajectories of most SADs can be fitted by one or two deceleration processes, while some special ones have positive accelerations during the descent. For the thermal properties, we selected 54 SADs, whose front and body could be clearly distinguished from the surrounding during the entire descent, to perform Differential Emission Measure analysis. It is revealed that the temperatures of the SAD front and body tend to increase during their downward courses, and the relationship between the density and temperature indicates that the heating is mainly caused by adiabatic compression.
\end{abstract}




\section{Introduction} \label{sec:intro}
Solar flares are rapid energy release phenomena in the solar atmosphere, which usually produce high-temperature flare loops in the corona. During a solar flare, we often see a dynamic region called the plasma sheet (also called fan or current sheet) above the flare loops, which is bright at extreme ultraviolet (EUV) and soft X-ray (SXR) wavelengths \citep[e.g.,][]{2013ApJ...766...39M,2018ApJ...866...29F,2019MNRAS.489.3183C}. In the plasma sheet, sometimes supra-arcade downflows \citep[SADs;][]{1999ApJ...519L..93M} are seen as tadpole-like structures leaving extended dark wakes during their descending motions toward the post-reconnection flare loops.

SADs are usually found in the early decay phases of long-term eruptive flares that produce coronal mass ejections (CME) \citep{1999ApJ...519L..93M, 2002ApJ...579..874S, 2003SoPh..217..247I, 2010ApJ...722..329S} and observed above the post-reconnection flare arcades with a face-on perspective, that is, the line of sight (LOS) is perpendicular to the arcades' axis. However, SADs sometimes can also appear in the impulsive phase of solar flares \citep{2004ApJ...605L..77A, 2007A&A...475..333K}. SADs have a typical lifetime of a few minutes, and their velocities are about 100--500 $\text{km}\ \text{s}^{-1}$. It is generally accepted that SADs are voids of plasma which are less dense and inherently lower in brightness than their surroundings \citep[e.g.,][]{2003SoPh..217..247I,2017ApJ...836...55R,2020ApJ...898...88X,2021ApJ...915..124L}. \citet{2014ApJ...786...95H} examined some SADs and concluded that they are cooler than the surrounding fan plasma through a Differential Emission Measure (DEM) analysis. In addition, \citet{2017A&A...606A..84C} found that the emission measure (EM) of SADs above 4 MK is smaller than that of the surrounding fan plasma. Recently, \citet{2022arXiv220301366Z} synthesized radio images of SADs through modeling and found that radio observations could help to obtain more accurate temperature and density information.

SADs are essentially caused by magnetic reconnection, which is a principal mechanism responsible for various types of solar activities. Additionally, \citet{2019PhRvL.123c5102S} found that SADs can generate the oscillation of upflows through vortex shedding. Studying the relationship between plasma heating and SADs can also help us understand the energy release during flares. \citet{2017ApJ...836...55R} found that the temperatures of regions with SADs tend to increase and the temperatures of regions without SADs decrease, and the heating mechanism was interpreted as adiabatic compression in front of SADs. A recent three-dimensional (3D) magnetohydrodynamic (MHD) simulation work about the plasma heating in the current sheet suggested that adiabatic compression is important for plasma heating in the late phase of the eruption \citep{2019ApJ...887..103R}. \citet{2020ApJ...898...88X} found that some plasma heating can be explained as adiabatic compression by SADs but others may be caused by different mechanisms. \citet{2021ApJ...915..124L} also suggested that the heating of SADs is caused by adiabatic compression and the viscous heating plays a limited role. Furthermore, \citet{2021Innov...200083S} reported a case that many SADs collide with the flare loops directly and found that these SADs can strongly heat the flare loops to 10--20 MK.

In the past, many observational \citep{1999ApJ...519L..93M, 2009ApJ...697.1569M, 2011ApJ...742...92W, 2014ApJ...786...95H, 2017A&A...606A..84C} and simulation \citep{2013ApJ...775L..14C,2014ApJ...796L..29G, 2014ApJ...796...27I,2015ApJ...807....6C, 2016ApJ...832...74Z} studies have been performed to investigate the formation mechanisms of SADs. \citet{1999ApJ...519L..93M} proposed that SADs are the cross-sections of evacuated flux tubes or wakes behind retracting loops \citep{ 2009ApJ...697.1569M, 2010ApJ...722..329S}. \citet{2011A&A...527L...5M} gave a different explanation that SADs are formed by intermittent, bursty magnetic reconnection in the plasma sheet or its upper coronal region \citep{2015ApJ...807....6C, 2016ApJ...832...74Z}.  \citet{2013ApJ...775L..14C} reported that SADs are outflow jets from reconnection sites penetrating into the denser flare arcades, which may explain the fact that SADs are not filled with the surrounding denser and hotter plasma immediately. In addition, \citet{2014ApJ...796L..29G} and \citet{2014ApJ...796...27I} interpreted SADs as a result of Rayleigh-Taylor instability (RTI) between the reconnection outflow and the denser plasma sheet. \citet{2020ApJ...898...88X} proposed that SADs are outflows of patchy and bursty magnetic reconnection because they can push surrounding plasma away and keep it from flowing back. Recently, \citet{2022NatAs.tmp...29S} proposed a new scenario that SADs are not embedded in the plasma sheet but located in the turbulent interface region between post--reconnection flare arcades and plasma sheet. And they further proposed that SADs are formed due to a mixture of RTI and Richtmyer-Meshkov instability (RMI).

However, statistical investigations of SADs are very rare, which hampers our understanding of both the nature and properties of SADs. \citet{2011ApJ...730...98S} performed a quantitative statistical study of SAD parameters like velocity, acceleration, area, magnetic flux, shrinkage energy, and reconnection rate, where up to 60 SADs were reported for a single flare event. However, the spatial and temporal resolutions of the data they uesd are relatively low. Based on data from the Atmospheric Imaging Assembly \citep[AIA;][]{2012SoPh..275...17L} on the  Solar Dynamics Observatory \citep[SDO;][]{2012SoPh..275....3P}, \citet{2021ApJ...915..124L} further studied 20 SADs in three flares with higher spatial and temporal resolutions. However, they did not investigate temporal evolutions of different properties of SADs in different phases of the flare. In this article, for the first time we used data from AIA to perform a statistical study of 81 SADs detected in a single flare event. Through identifying downflow trajectories of all the SADs and performing DEM analyses of 54 ones of these SADs, we studied their kinematic and thermal properties, respectively, and further investigated the evolution of these properties during the flare. The remainder of this paper is organized as follows. Section \ref{sec:obs} presents the AIA observations used in this work. Section \ref{sec:Kinematics} presents our analysis results of kinematic properties of SADs. Section \ref{sec:Thermal} shows the thermal properties of SADs and analysis results about the heating of SADs. In Section \ref{sec:Con}, we briefly summarize our finding.

\section{Observation} \label{sec:obs}
The data used in our study came from the SDO/AIA. The AIA is convenient to observe flares on the solar limb because it monitors the entire Sun up to a radial distance of $\sim$1.30 $R_{\odot}$ from the disk center \citep{2012SoPh..275...17L}. Because the 335, 94, and 131 {\AA} channels of AIA are sensitive to plasma with high temperatures of $\sim$3--10 MK \citep{2012SoPh..275...41B}, we can easily identify SADs in these channels \citep{2011ApJ...727L..52R}. These observations were taken at a high temporal cadence of 12 s and angular resolution of $\sim$0.6\arcsec. The SADs studied here were identified from an M5.0 flare erupting on the northwest limb of the Sun starting at about 13:08 UT on 2013 May 22, which was also reported by \citet{2018ApJ...866...29F}. The delay phase of the flare lasted for about 3 hours, and the flare plasma sheet was nearly perpendicular to the LOS from the Earth. So we could easily identify 81 SADs in this flare event. The AIA data were prepared using the {\tt aia\_prep} routine, available in the Solar Software \citep[SSW;][]{1998SoPh..182..497F} package. This routine de-rotates the AIA images from the four telescopes, aligns them, and re-scales the images to achieve the same plate scale.

\section{Kinematic properties} \label{sec:Kinematics}
\subsection{Identification of SADs} \label{subsec:Identification}
SADs usually first appear as dark structures at high altitudes above the solar surface and then move downward to the flare arcades. We often observe SADs in high-temperature passbands like AIA 131 {\AA}. In this flare, we tracked the trajectories of SADs frame by frame with the AIA image sequences of the 131 {\AA}, 94 {\AA}, and 335 {\AA} passbands, and finally identified 81 obvious SADs. In Table \ref{tab:SADs}, we list the details of each identified SAD. All of the SADs can be found in the AIA 131 {\AA} channel, but only the SADs in the late decay phase can be found in the AIA 94 {\AA} or 335 {\AA} channels. It is worth noting here that none of the SADs could be observed at their origin sites and all of them are actually observed well below their true initial height. And the observed ``initial height" (hereafter we refer to ``appearance height") is determined by the instrument-dependent detection (e.g., wavelength, dynamic range, and field of view) and the evolution of plasma environment above the flare arcade \citep{2011ApJ...730...98S}.

\begin{figure*}
\centering
\includegraphics [width=0.75\textwidth]{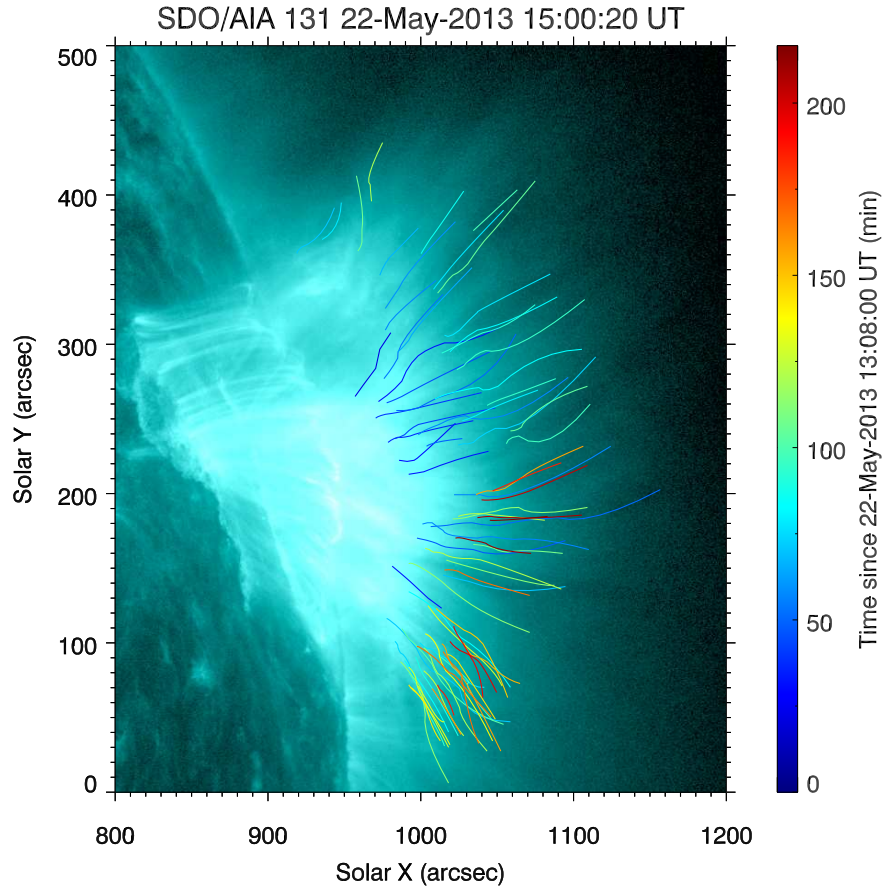}
\caption{An SDO/AIA 131 {\AA} image showing the hot flare region overlaid with trajectories of the 81 SADs. The curves with different colors represent the trajectories of SADs appearing at different times. An animation of AIA 131 {\AA} observations covering the time from 13:08 UT to 16:55 UT on 2013 May 22 is available as supplementary material.
}
\label{fig:SADs}
\end{figure*}

In Figure \ref{fig:SADs}, we plot trajectories of all the 81 SADs on one 131 {\AA} image with different colors according to their appearance times. It could be found that, as the flare evolved, the appearance heights of SADs become larger especially for the SADs in the north part (see the associated animation), which possibly indicate the lift of the bottom of the current sheet shown in \citet{2012MNRAS.425.2824M}. However, as mentioned in \citet{2011ApJ...730...98S}, this trend might also be caused by the fact that more hot plasma is produced in higher places above the flare arcade as the flare progresses, which could provide a brighter background and thus a favor to observe more SADs at increasing heights. At the same time, in the decay phase of the flare, the region where SADs concentrated shifts from the north side to the south side. Combining the observations from the Large Angle and Spectrometric COronagraph (LASCO) C2, we found that there was a secondary eruption in the south part of the plasma sheet, which likely results in this shift.

\subsection{Measurements of velocity and acceleration for SADs with a constant acceleration} \label{subsec:Measurements}
\begin{figure*}
\centering
\includegraphics[width=0.99\textwidth]{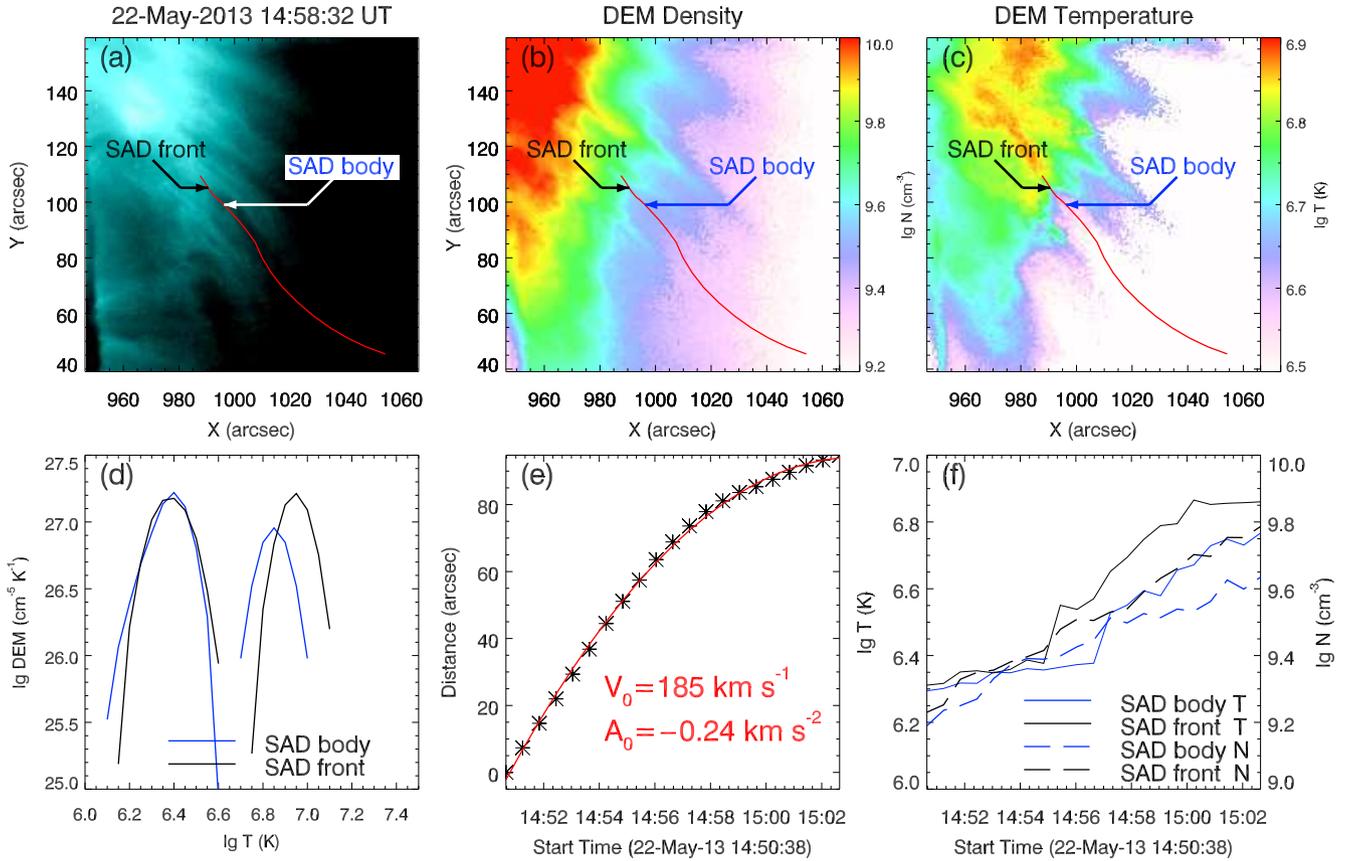}
\caption{Analysis results of a typical SAD. (a): The trajectory of a SAD (red curve) marked on an AIA 131 {\AA} image. (b): The corresponding density map obtained through the DEM analysis. (c): The corresponding temperature map obtained through the DEM analysis. The front and body sites of an SAD are marked with arrows in panels (a)-(c); (d): The DEM curves in the two sites; (e): Fitting of the trajectory of the SAD. $\text{V}_{0}$ and $\text{A}_{0}$ represent the initial velocities and acceleration, respectively. (f): Temperature and density variations of the front and body regions of this SAD.
}
\label{fig:case1}
\end{figure*}

After determining the trajectories of SADs, we further analyzed the kinematics of each SAD. Figure \ref{fig:case1} shows the typical analysis process of a SAD that we employed in the present work. The red curve indicates the trajectory of the SAD, and the front and body of the SAD at 14:58:32 UT are marked with the black and blue arrows in panel (a), respectively. The star marks in panel (e) represent the moving distance of this SAD along its trajectory at different times. It is revealed that this SAD undergoes a significant deceleration and finally disappears at a low altitude. The deceleration of SAD was frequently reported in previous observations \citep[e.g.,][]{2021Innov...200083S,2020ApJ...898...88X,2021ApJ...915..124L}, which could be caused by the drag force \citep{2006ApJ...642.1177L,2013ApJ...776...54S}. \citet{2022NatAs.tmp...29S} further proposed that maybe the interface region between the reconnection current sheet and the flare arcades causes the outflows to abruptly slow down.

To obtain the projective velocity and acceleration of this SAD, we used one 2nd order polynomial fitting to its trajectory as shown in Figure \ref{fig:case1}(e). $\text{V}_{0}$ and $\text{A}_{0}$ represent the initial velocity and acceleration of the SAD, respectively. We found that the SAD initially descends with a velocity of 185 $\text{km}\ \text{s}^{-1}$ but decelerates rapidly with an acceleration of --0.24 $\text{km}\ \text{s}^{-2}$ (here the minus sign means deceleration), and eventually disappears with a small velocity. It is worth noting that among the 81 SADs studied here, there are 31 ones ($\sim38\%$) with the trajectory that could be approximated by one 2nd order polynomial fitting. The others SADs have varying accelerations during the entire descent (see more details in Section \ref{subsec:special}).

\subsection{Measurements of velocity and acceleration for SADs with varying accelerations} \label{subsec:special}
\begin{figure*}
\centering
\includegraphics[width=0.99\textwidth]{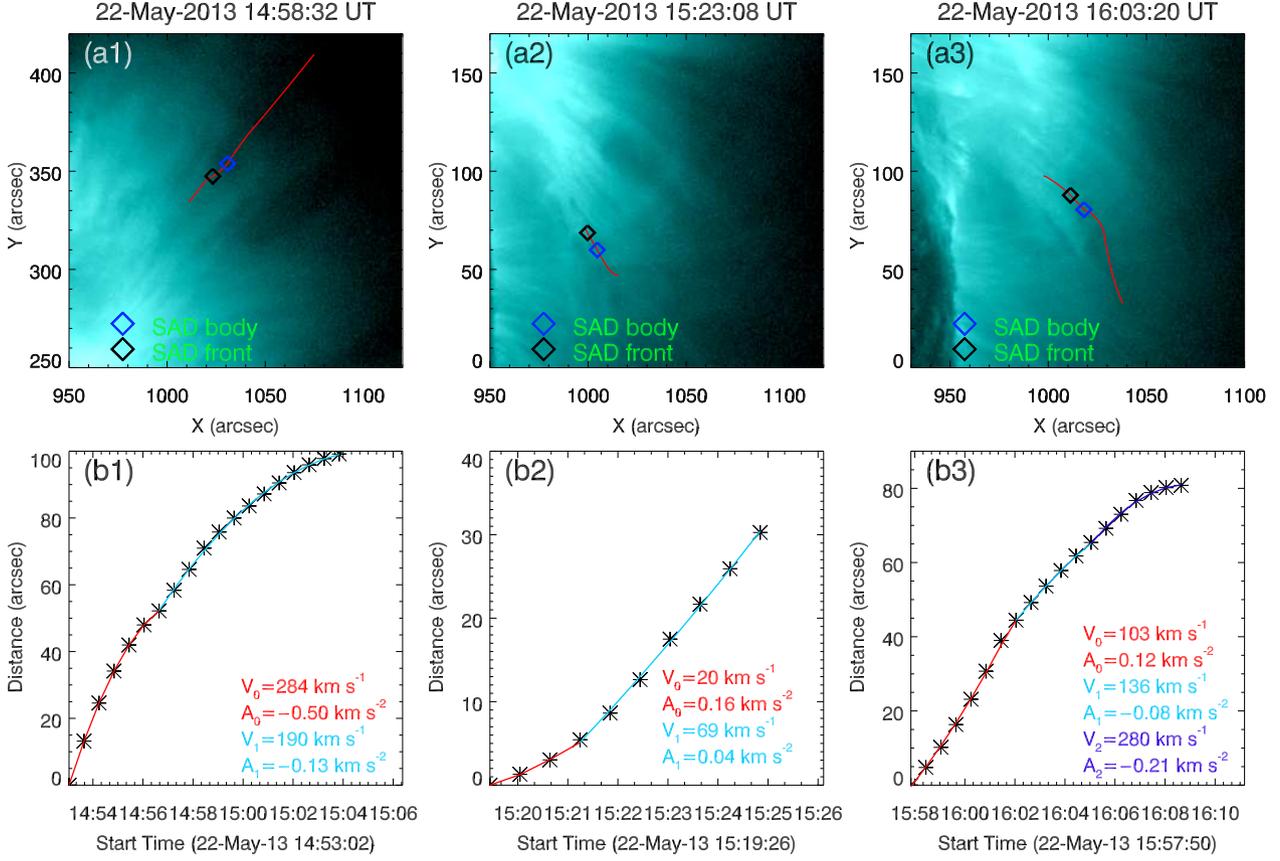}
\caption{Three typical events of SADs with varying accelerations. (a1)-(a3): Trajectories of these SADs (red curves) marked on an AIA 131 {\AA} image. (b1)-(b3): Corresponding fitting results of these trajectories. The results of the multi-segment fit are shown in different colors.
}
\label{fig:specialsads}
\end{figure*}

Besides the 31 SADs experiencing only one stage of deceleration mentioned above, the remaining SADs have varying accelerations. Among them, 28 SADs ($\sim35\%$) undergo two decelerations with different accelerations, and 22 SADs ($\sim27\%$) have more complex kinetic characteristics, which experience at least one acceleration. In Figure \ref{fig:specialsads}, we show three typical events of these SADs with varying accelerations and the corresponding fitting results.

\begin{figure*}
\centering
\includegraphics[width=0.75\textwidth]{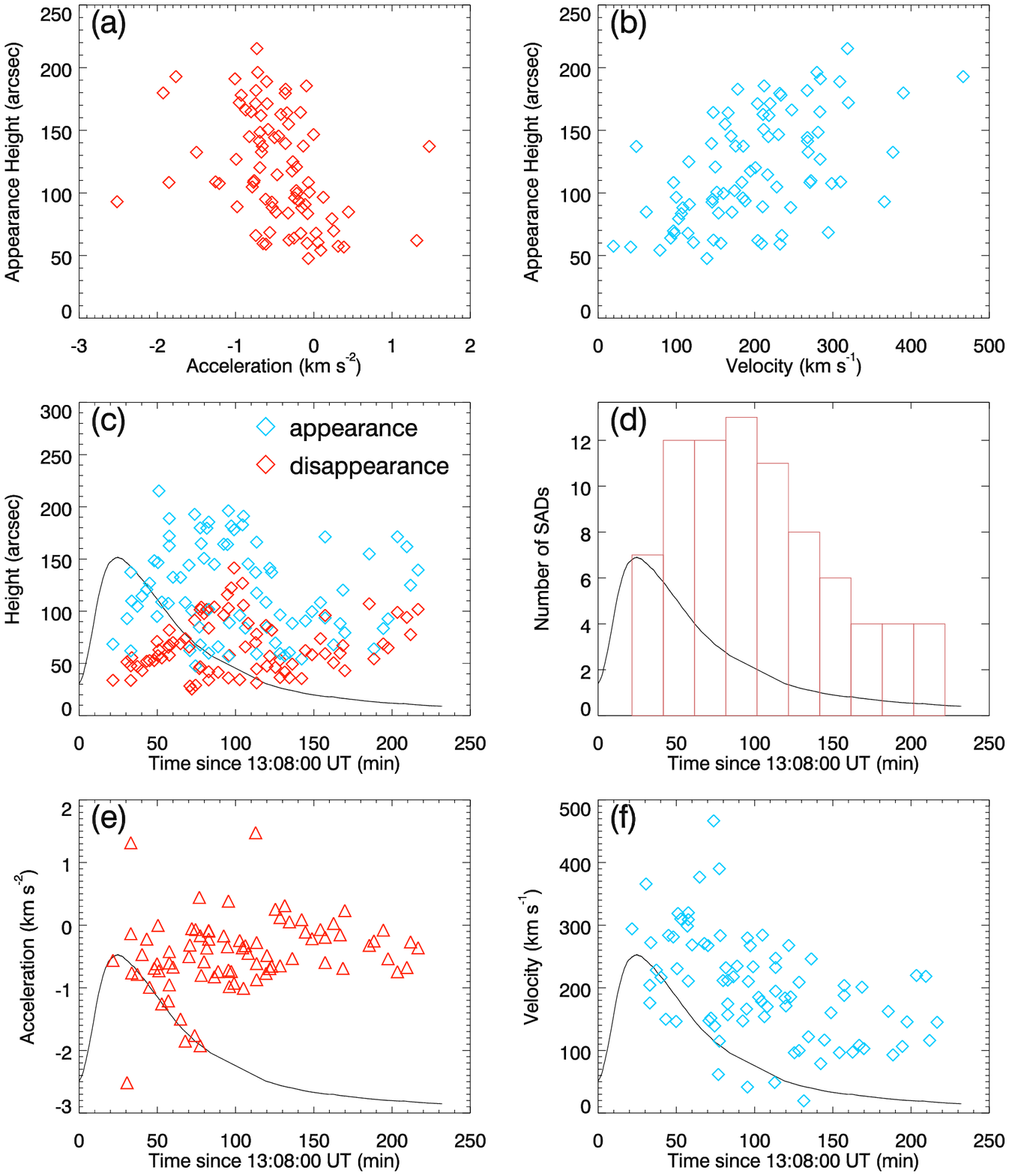}
\caption{Statistical result of the kinematic parameters of the 81 SADs. (a): Scatter plot of the initial acceleration vs appearance height. (b): Scatter plot of the initial velocity vs appearance height. (c): Scatter plot of the appearance time vs appearance and disappearance height. (d): Occurring frequency of SADs as a function of time. (e): Scatter plot of the appearance time vs acceleration. (f): Scatter plot of the appearance time vs appearance velocity. The black curves in panels (c)-(f) represent the GOES SXR 1--8 {\AA} flux variation.
}
\label{fig:plot_dynpara}
\end{figure*}

For the SAD in Figures \ref{fig:specialsads}(a1) and (b1), it has an initial velocity of 284 $\text{km}\ \text{s}^{-1}$ and then experience two stages of deceleration with accelerations of --0.5 and --0.13 $\text{km}\ \text{s}^{-2}$, respectively. The SAD shown in panels (a2) and (b2) appears with a velocity of 20 $\text{km}\ \text{s}^{-1}$ and experiences two stages of acceleration with accelerations of 0.16 and 0.04 $\text{km}\ \text{s}^{-2}$, respectively. The SAD shown in panels (a3) and (b3) appears with an initial velocity of 103 $\text{km}\ \text{s}^{-1}$, and then undergoes one acceleration and two decelerations with accelerations of 0.12, --0.08, --0.21 $\text{km}\ \text{s}^{-2}$, respectively. We speculate that it is the collision or merging between SADs and a turbulent environment that affect the motion of SADs.

\subsection{Statistical results of kinematic properties} \label{subsec:sum}
From the analysis mentioned above, we obtained kinematic parameters including the height, projective velocity, and acceleration of all the 81 SADs (see details in Table \ref{tab:SADs}). In this work, the height refers to the distance between the SAD and the solar limb. And the appearance height ($\text{H}_{app}$) and disappearance height ($\text{H}_{dis}$) of each SAD are calculated. Figure \ref{fig:plot_dynpara} shows scatter plots for the 81 SADs. Panel (b) shows that the appearance height of SADs has a positive correlation with the initial projective velocity. Possibly, due to the low emission contrast between SADs and the surroundings at the beginning, SADs with lower appearance heights could have already experienced a certain deceleration process before they are identified. As a result, such SADs are associated with smaller initial velocities when they are identified. Essentially, it is affected by the difficult determination of ``initial height" of SADs as discussed in Section \ref{subsec:Identification}.

As shown in Figure \ref{fig:plot_dynpara}(c), the appearance heights of the SADs are about 50--200 Mm, and the disappearance heights of the SADs are about 20--130 Mm. Both of the appearance and disappearance heights increase in the first half period of our observation, which can also be clearly seen from Figure \ref{fig:SADs}. Panel (d) shows that the SADs mainly appear in the delay phase of the flare. As shown in panels (e) and (f), the initial velocity of SADs is about 100--400 $\text{km}\ \text{s}^{-1}$, and the acceleration is about -2--0 $\text{km}\ \text{s}^{-2}$. The initial velocity range is similar to that reported by \citet{2011ApJ...730...98S}.

\section{Thermal properties} \label{sec:Thermal}
To obtain the accurate thermal properties of SADs, we selected 54 SADs, whose front and body could be clearly distinguished from the surrounding during the entire lifetime, to perform a DEM analysis using images in six AIA channels: 94 {\AA}, 131 {\AA}, 171 {\AA}, 193 {\AA}, 211 {\AA}, and 335 {\AA}. To increase the signal-to-noise ratio, we averaged three adjacent frames to achieve a cadence of 36 s. The codes we used to perform the DEM analysis were written by \citet{2015ApJ...807..143C} and improved by \citet{2018ApJ...856L..17S}. This method obtains DEM solutions after we input narrow-band EUV images from different AIA channels and the temperature response functions of these channels. By using this code, the DEM results derived from AIA images alone are much more consistent with high--temperature (X-ray) observations than other methods \citep{2018ApJ...856L..17S}. We then calculated the DEM weighted average temperature ($T_{dem}$), given by
\begin{equation}
\label{t_em.eq}
T_{dem} = \frac{\int T \times DEM(T)dT}{\int DEM(T)dT}
\end{equation}
and the total emission measure ($EM_{tot}$), given by
\begin{equation}
\label{eq:em}
EM_{tot} =\int DEM(T)dT.
\end{equation}
We obtained an EM by integrating the DEM over the temperature range from lg T/K=6.0 to lg T/K=7.5, which includes a low--temperature component and a high--temperature component (e.g., Figure \ref{fig:case1}(d)). The low--temperature component always peaks around 2 MK, because it comes from the background coronal plasma. The high-temperature component often peaks at 7--10 MK, which mainly comes from the emission of SADs. We further calculated the electron density using equation:
\begin{equation}
EM = N^{2}L
\end{equation}
where $N$ is the electron density. We assumed a characteristic path length $L=10^9$ cm \citep{2017ApJ...836...55R}.

From the DEM analysis, we obtained the temperature and density maps (Figures \ref{fig:case1}(b),(c)). The temperature and density at the front and body of each SAD at each time step were also calculated. It has been suggested that the heating of the SAD front may be caused by adiabatic compression \citep{2017ApJ...836...55R}. Here, we further explore this issue statistically using a method that is similar to that used by \citet{2021ApJ...915..124L}.

\begin{figure*}
\centering
\includegraphics[width=0.99\textwidth]{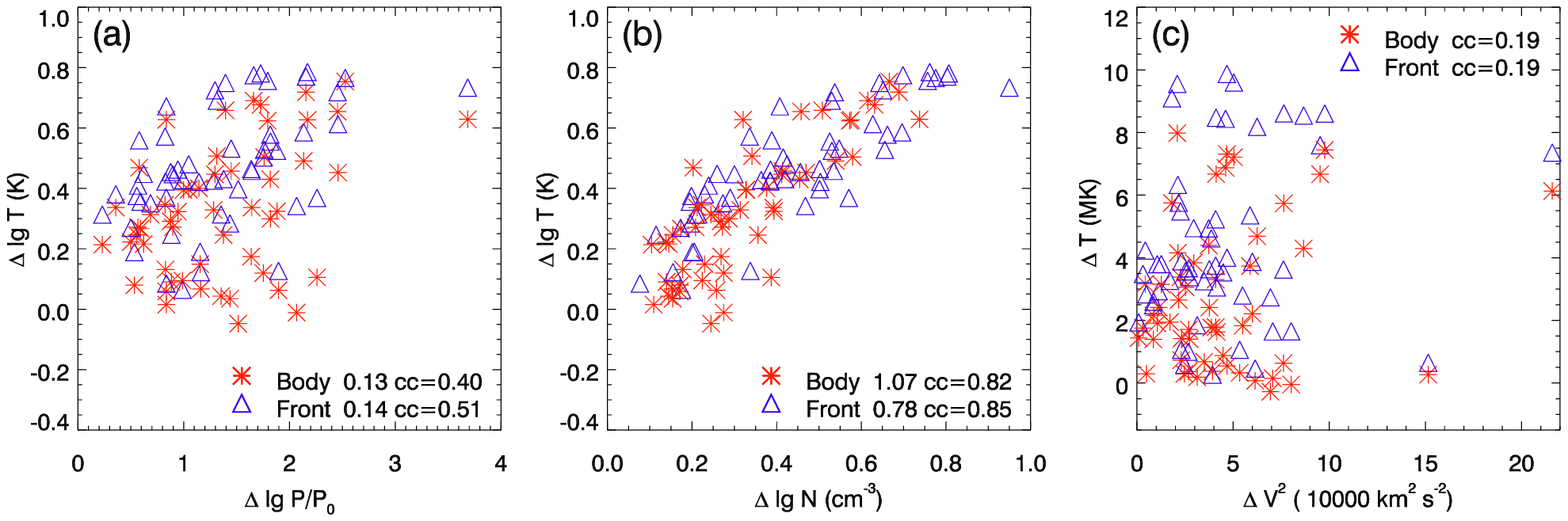}
\caption{(a): Scatter plot of the variation of pressure vs that of temperature. (b): Scatter plot of the variation of density vs that of temperature. (c): Scatter plot of the variation of velocity squared vs that of temperature. The blue triangles and red stars represent the fronts and bodies of SADs;
}
\label{fig:plot_adiabatic}
\end{figure*}

\begin{figure*}
\centering
\includegraphics[width=0.85\textwidth]{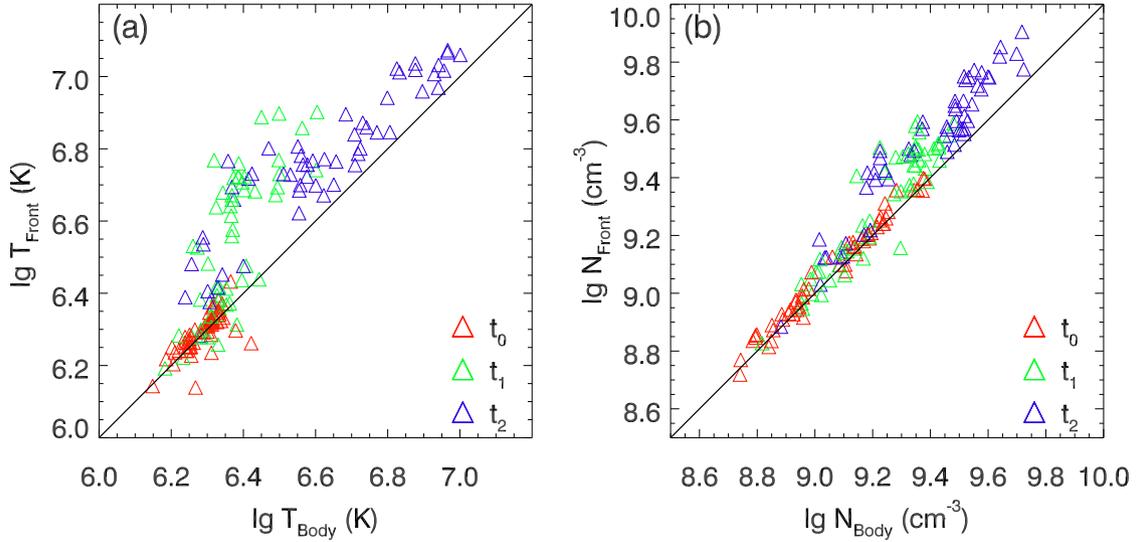}
\caption{Temperature and density comparisons between the fronts and bodies of 54 SADs. (a): Scatter plot of the body and front temperatures at three different time points (start $\text{t}_{0}$, middle $\text{t}_{1}$, and end time $\text{t}_{2}$). (b): Scatter plot of the body and front densities. The red, green, and blue triangles represent the three different stages of the SAD trajectories.
}
\label{fig:plot_compareTEM}
\end{figure*}

For an ideal gas, an adiabatic process means that the temperature and pressure obey the following equation,
\begin{equation}
\centering
\frac{P^{\gamma-1}}{T^{\gamma}} = C1,
\label{eq:adiabatic1}
\end{equation}
where $T$ is the plasma temperature, $P$ is the plasma pressure, $\gamma = 5/3$ is the polytropic index and $C1$ is a constant. Then we use the ideal gas law ( $P=KNT$ ) to derive the following equation:
\begin{equation}
\centering
\frac{N^{\gamma-1}}{T} = C2
\label{eq:adiabatic2}
\end{equation}
where $C2$ is a constant. Because the deceleration of SADs is relatively small, we assumed that the atmosphere above post-reconnection flare loops is in static equilibrium. Thus, the pressure decays exponentially with the height as:
\begin{equation}
\centering
P = P_{0} e^{-H/H_{S}}
\label{eq:pressure}
\end{equation}
where $P_{0}$ is a constant, $H_{S}$ is the average scale height at $T$ = 1 MK. The specific value of $H_{S}$ has no effect on the following analysis. Here, we use the heights of the SADs projected to solar limb instead of the actual heights of the SADs above the solar limb. According to Equations \ref{eq:adiabatic1} and \ref{eq:adiabatic2}, we produced scatter plots for the relationship between pressure and temperature variations (Figure \ref{fig:plot_adiabatic}(a)) and scatter plots for the relationship between density and temperature variations (Figure \ref{fig:plot_adiabatic}(b)) for both the SAD front and body. Compared to \citet{2021ApJ...915..124L}, here we further performed the adiabatic correlation analysis of temperature and density. And the following parameters were calculated: $\Delta (lg T)$, $\Delta (lg N)$, and $\Delta (lg P/P_{0})$. For each SAD, the variations were estimated by subtracting the value at the first moment from that at the last moment.

As shown in Figure \ref{fig:plot_adiabatic}(a), we did not obtain a good correlation between the pressure and temperature in logarithmic scale as in \citet{2021ApJ...915..124L}, although we did a similar analysis. A possible reason is that the pressure we calculated according to Equation \ref{eq:pressure} is subject to a large uncertainty because the heights of SADs we used here are the projected heights instead of the actual ones. This means that in different events, the different LOS inclinations could affect the values of projective heights and thus the estimated values of pressure. As a result, the relationship between the pressure and temperature could change.

Figure \ref{fig:plot_adiabatic}(b) reveals that the slopes for the $\Delta (lg T)$ -- $\Delta (lg N)$ relationship are about 0.78 and 1.07 for the SAD front and body, respectively. These values are close to 0.67 ($N \sim T^{2/3}$ from equation \ref{eq:adiabatic2}), suggesting that adiabatic compression plays an important role in heating the plasma at the front and in the body of SADs. The correlation coefficients between $\Delta (lg T)$ and $\Delta (lg N)$ are about 0.85 and 0.82 for the SAD front and body, respectively. As a result, we think that both the front and body of SADs may have experienced the adiabatic compression process, which could be the critical mechanism of heating SADs when they flow downwards the loop top, as previously claimed by \citet{2017ApJ...836...55R} and \citet{2020ApJ...898...88X}. It is worth noting that in the actual situation the characteristic path length $L$ could be slightly various at different times, which thus would affect the estimated density values ($N$) of SADs as well as the relationship between the density and temperature variations.

We also examined the possibility that the kinetic energy of SADs may be converted into thermal energy through viscous dissipation. For this we investigated the statistical relationship between $\Delta V^2$ and $\Delta T$, which are proxies of the variations of kinetic energy and thermal energy, respectively. For the \textcolor[rgb]{0.5,0.,0.}{54} SADs, the result is shown in Figure \ref{fig:plot_adiabatic}(c). We can see that the correlation coefficient is not large enough either for the SAD front or the SAD body, which is different from the results obtained by \citet{2021ApJ...915..124L}. This result indicates that the dissipation of the kinetic energy of SADs is not dominant (or very limited) in the entire heating process of the SAD front or body during the flare. Note that the velocity we derived here is only the projected component of the three-dimensional velocity in the plane of sky, which could also affect our calculation of the correlation between the changes of kinetic energy and thermal energy.

In Figure \ref{fig:plot_compareTEM}, we compared the temperature and density of the front and body during the downward motions of SADs. The thermal properties of SAD in three time points (start time $\text{t}_{0}$, middle time $\text{t}_{1}$, and end time $\text{t}_{2}$) are shown with different colors, respectively. It is obvious that the temperature of the front is usually higher than that of the body (panel (a)), which is similar to the results reported by \citet{2021ApJ...915..124L}. We also found that some SADs have higher body temperatures compared to their fronts. A similar case was also studied by \citet{2014ApJ...786...95H}, where the SAD was hotter than surrounding plasma. The front density is also higher than the body density in most SADs (panel (b)), which is similar to the result of \citet{2014ApJ...786...95H}. In addition, the temperatures of the SAD front and body tend to increase during the downward course of SADs, which are consistent with Figure \ref{fig:case1}(f) and can be explain by adiabatic compression heating occurring in the entire process.

\section{Summary} \label{sec:Con}
In this work, we identified 81 SADs in one flare event with AIA observations. This is the largest sample of SADs ever obtained in a single flare, which have statistical investigation of the kinematic and thermal properties simultaneously. We extracted the trajectories of the SADs from space-time diagrams. All of these SADs can be found in the AIA 131 {\AA} band, and some of them can also be identified in the AIA 94 {\AA} and 335 {\AA} bands in the late decay phase. The appearance heights of the SADs become larger with time especially in the north part, possibly indicating the lift of the bottom of the current sheet during the flare or the evolution of plasma environment above the flare arcade. In addition, there is a shift of the SAD concentration region from the north part to the south part, which is possibly related to a secondary eruption in the south side. The trajectories of most SADs can be fitted by one (31 SADs) or two (28 SADs) deceleration processes. Meanwhile, some special ones (22 SADs) have positive accelerations during the descent, which could be caused by collision or merging between SADs and a turbulent environment.

We found that the initial velocities of SADs are about 100--400 $\text{km}\ \text{s}^{-1}$, and the accelerations are about -2--0 $\text{km}\ \text{s}^{-2}$. These are similar to the previous results in \citet{2011ApJ...730...98S}. The significant deceleration in most SADs are also be modeled in \citet{2022NatAs.tmp...29S} and are attributed to the existence of turbulence interface region. It is also found that the appearance heights of SADs have a positive correlation with the initial velocities, which is not reported in previous studies due possibly to the low spatial resolution of data. Additionally, the initial velocities are lower than the typical coronal Alfv$\acute{e}$n velocity, which means that these SADs are possibly not reconnection outflows themselves. The model of RTI-generated SADs proposed by \citet{2014ApJ...796L..29G} could explain the low velocity. But it is worth noting that because the SADs are projectively detected at the appearance height where they have experienced a deceleration process, the initial velocities calculated here should be the lower limit value of the true speed of the SADs when they are formed \citep{2020AA...642A..44H}.

We have carried a DEM analysis for 54 SADs, whose front and body could be clearly distinguished from the surrounding during the entire descent. The DEM of the SAD body at the temperature range of $10^{6.0-7.5}$ K is generally smaller than that of the SAD front region (e.g., Figure \ref{fig:case1}(d)), which is consistent with previous findings \citep{2012ApJ...747L..40S,2014ApJ...786...95H}. We found that the temperature of SAD front increases when an SAD passes by, which is consistent with the results of \citet{2017ApJ...836...55R} and \citet{2020ApJ...898...88X}. These results suggest that SADs are heated up when falling. By using an analysis method similar to that of \citet{2021ApJ...915..124L}, we further demonstrated that both the front and body of SADs evolve adiabatically, confirming that adiabatic compression could be the critical mechanism of the entire heating process \citep{2017ApJ...836...55R,2020ApJ...898...88X,2021ApJ...915..124L}. We also noticed that the correlation coefficient between the kinetic energy and thermal energy is not large for both the front and body of SADs. It suggests that the dissipation of the kinetic energy of SADs is not dominant in heating SADs during the flare.

\section*{Acknowledgements}
The authors are cordially grateful to the anonymous referee for his/her constructive comments and suggestions improving this paper. This work is supported by the National Key R\&D Program of China No. 2021YFA0718600, NSFC grants 11825301 and 11903050, the Strategic Priority Research Program of the Chinese Academy of Sciences (XDB41000000), and the NAOC Nebula Talents Program. The data used here are courtesy of the \emph{SDO} and \emph{GOES} science teams. \emph{SDO} is a mission of NASA's Living With a Star Program. We thank Prof. Bin Chen and Dr. Sijie Yu for helpful discussion.

\section*{Data availability}
The data underlying this article will be shared on reasonable request to the corresponding author.



\bibliographystyle{mnras}
\bibliography{SADs_paper} 

\onecolumn
\begin{longtable}{cccccccccc}
\caption{Physical parameters of the 81 SADs. \label{tab:SADs}}\\
\hline\hline\noalign{\smallskip}
SADs & Start time & End time & $\text{V}_{0}$  & H$_{app}$ & H$_{dis}$ & A$_{0}$    & A$_{1}$    & A$_{2}$   & For DEM \\
 & & & $(\text{km}\ \text{s}^{-1})$ & $(\text{Mm})$ & $(\text{Mm})$ & $(\text{km}\ \text{s}^{-2})$ & $(\text{km}\ \text{s}^{-2})$ & $(\text{km}\ \text{s}^{-2})$ & \\
\hline\noalign{\smallskip}
1  & 13:29:38   & 13:32:02 & 294 & 50     & 25     & --0.56 & /  & /  & \checkmark\\
2  & 13:38:38   & 13:42:14 & 365 & 68     & 37     & --2.51 & 0.81  & /  &         \\
3  & 13:41:02   & 13:44:38 & 204 & 45     & 25     & 1.31  & 0.76  & /  &         \\
4  & 13:41:02   & 13:48:50 & 175 & 100    & 35     & --0.13 & /  & /  & \checkmark\\
5  & 13:41:38   & 13:45:14 & 272 & 80     & 40     & --0.76 & /  & /  & \checkmark\\
6  & 13:45:14   & 13:52:26 & 228 & 76     & 34     & --0.79 & --0.08 & /  & \checkmark\\
7  & 13:48:14   & 13:55:26 & 216 & 83     & 31     & --0.47 & /  & /  & \checkmark\\
8  & 13:51:14   & 13:59:38 & 149 & 88     & 38     & --0.22 & /  & /  & \checkmark\\
9  & 13:53:02   & 14:00:14 & 283 & 92     & 38     & --0.99 & --0.12 & /  & \checkmark\\
10 & 13:56:02   & 14:07:26 & 281 & 108    & 38     & --0.69 & --0.18 & /  &         \\
11 & 13:57:50   & 14:01:26 & 146 & 69     & 52     & --0.62 & /  & /  & \checkmark\\
12 & 13:58:26   & 14:05:38 & 230 & 106    & 42     & 0.00  & --0.07 & /  & \checkmark\\
13 & 13:59:02   & 14:11:38 & 318 & 156    & 46     & --0.73 & --0.14 & /  & \checkmark\\
14 & 14:00:50   & 14:06:14 & 309 & 79     & 40     & --1.25 & --0.10 & /  & \checkmark\\
15 & 14:05:02   & 14:08:02 & 298 & 78     & 48     & --1.20 & /  & /  &         \\
16 & 14:05:38   & 14:14:02 & 308 & 137    & 59     & --0.60 & /  & /  &         \\
17 & 14:05:38   & 14:14:38 & 319 & 125    & 49     & --0.95 & --0.20 & /  & \checkmark\\
18 & 14:05:38   & 14:19:26 & 210 & 118    & 42     & --0.42 & --0.11 & /  &         \\
19 & 14:08:02   & 14:13:26 & 268 & 96     & 51     & --0.67 & /  & /  & \checkmark\\
20 & 14:12:50   & 14:18:14 & 376 & 96     & 50     & --1.49 & --0.46 & /  &         \\
21 & 14:15:50   & 14:20:02 & 270 & 79     & 54     & --1.85 & 0.39  & /  & \checkmark\\
22 & 14:18:14   & 14:23:38 & 267 & 105    & 48     & --0.50 & /  & /  &         \\
23 & 14:18:50   & 14:24:14 & 147 & 45     & 21     & --0.32 & --0.61 & /  & \checkmark\\
24 & 14:20:02   & 14:32:38 & 151 & 73     & 18     & 0.06  & --0.28 & /  & \checkmark\\
25 & 14:21:50   & 14:29:02 & 466 & 140    & 66     & --1.76 & --0.45 & /  & \checkmark\\
26 & 14:22:26   & 14:31:26 & 139 & 35     & 21     & 0.07  & --0.18 & /  &         \\
27 & 14:24:50   & 14:30:50 & 61  & 62     & 33     & 0.44  & --0.41 & /  &         \\
28 & 14:25:26   & 14:29:02 & 114 & 49     & 34     & --0.17 & /  & /  &         \\
29 & 14:25:26   & 14:32:02 & 390 & 131    & 74     & --1.92 & --0.97 & /  & \checkmark\\
30 & 14:26:02   & 14:31:26 & 283 & 120    & 75     & --0.80 & /  & /  &         \\
31 & 14:27:50   & 14:33:14 & 211 & 109    & 71     & --0.58 & /  & /  &         \\
32 & 14:29:38   & 14:35:02 & 232 & 130    & 74     & --0.36 & /  & /  & \checkmark\\
33 & 14:30:50   & 14:36:14 & 156 & 43     & 25     & 0.08  & --0.70 & /  & \checkmark\\
34 & 14:30:50   & 14:41:38 & 212 & 135    & 61     & 0.09  & --0.02 & /  & \checkmark\\
35 & 14:30:50   & 14:41:38 & 174 & 74     & 30     & --0.22 & /  & /  & \checkmark\\
36 & 14:34:26   & 14:39:50 & 217 & 105    & 76     & --0.82 & 0.20  & /  & \checkmark\\
37 & 14:36:50   & 14:44:02 & 234 & 48     & 30     & --0.74 & --0.32 & /  & \checkmark\\
38 & 14:40:26   & 14:48:14 & 147 & 119    & 70     & --0.17 & /  & /  & \checkmark\\
39 & 14:42:50   & 14:48:14 & 166 & 119    & 84     & --0.34 & /  & /  & \checkmark\\
40 & 14:43:26   & 14:52:26 & 41  & 41     & 26     & 0.38  & --0.23 & /  & \checkmark\\
41 & 14:43:26   & 14:53:02 & 279 & 142    & 75     & --0.72 & --0.15 & /  & \checkmark\\
42 & 14:44:02   & 14:51:14 & 210 & 65     & 42     & --0.98 & --0.21 & /  &         \\
43 & 14:45:14   & 14:50:38 & 267 & 132    & 89     & --0.74 & /  & /  & \checkmark\\
44 & 14:47:02   & 14:50:38 & 233 & 129    & 103    & --0.93 & /  & /  & \checkmark\\
45 & 14:50:38   & 15:02:38 & 185 & 70     & 25     & --0.24 & /  & /  &         \\
46 & 14:52:26   & 14:59:38 & 178 & 133    & 92     & --0.36 & /  & /  & \checkmark\\
47 & 14:53:02   & 15:03:50 & 284 & 139    & 77     & --1.00 & --0.26 & /  & \checkmark\\
48 & 14:54:14   & 15:01:26 & 153 & 61     & 48     & --0.33 & /  & /  &         \\
49 & 14:56:02   & 15:05:02 & 169 & 106    & 64     & --0.45 & --0.07 & /  & \checkmark\\
50 & 15:00:50   & 15:06:14 & 48  & 100    & 51     & 1.47  & --1.89 & /  &         \\
51 & 15:01:26   & 15:11:38 & 194 & 85     & 32     & --0.28 & /  & /  & \checkmark\\
52 & 15:01:26   & 15:11:38 & 232 & 43     & 23     & --0.61 & --0.19 & /  &         \\
53 & 15:01:26   & 15:12:14 & 247 & 121    & 57     & --0.87 & --0.11 & /  & \checkmark\\
54 & 15:07:26   & 15:12:50 & 183 & 79     & 63     & --0.77 & 0.16  & /  & \checkmark\\
55 & 15:08:02   & 15:18:50 & 171 & 61     & 34     & --0.48 & --0.14 & /  & \checkmark\\
56 & 15:09:50   & 15:21:14 & 267 & 103    & 41     & --0.69 & --0.06 & /  &         \\
57 & 15:11:02   & 15:20:02 & 185 & 100    & 59     & --0.66 & 0.05  & /  &         \\
58 & 15:13:26   & 15:18:50 & 96  & 51     & 34     & 0.25  & --0.33 & /  & \checkmark\\
59 & 15:16:26   & 15:23:38 & 208 & 43     & 26     & --0.65 & --0.16 & /  & \checkmark\\
60 & 15:16:26   & 15:25:26 & 99  & 70     & 39     & 0.12  & --0.04 & /  & \checkmark\\
61 & 15:19:26   & 15:24:50 & 19  & 42     & 30     & 0.31  & 0.07  & /  &         \\
62 & 15:22:26   & 15:27:50 & 121 & 44     & 26     & 0.05  & --0.61 & /  &         \\
63 & 15:24:14   & 15:33:14 & 246 & 64     & 36     & --0.53 & --0.23 & /  & \checkmark\\
64 & 15:30:14   & 15:37:26 & 79  & 39     & 26     & 0.09  & --0.63 & /  &         \\
65 & 15:32:38   & 15:39:50 & 116 & 66     & 45     & --0.11 & /  & /  &         \\
66 & 15:36:50   & 15:47:38 & 160 & 72     & 42     & --0.22 & /  & /  & \checkmark\\
67 & 15:42:14   & 15:49:26 & 96  & 79     & 54     & 0.07  & /  & /  & \checkmark\\
68 & 15:45:14   & 15:54:14 & 187 & 68     & 43     & --0.19 & --0.35 & /  & \checkmark\\
69 & 15:45:14   & 15:54:50 & 203 & 124    & 70     & --0.60 & --0.14 & /  & \checkmark\\
70 & 15:50:38   & 15:56:02 & 97  & 49     & 37     & 0.03  & --0.72 & /  &         \\
71 & 15:54:50   & 16:03:50 & 108 & 64     & 43     & --0.15 & 0.04  & --0.18 & \checkmark\\
72 & 15:56:38   & 16:03:50 & 201 & 87     & 49     & --0.69 & --0.17 & /  & \checkmark\\
73 & 15:57:50   & 16:08:38 & 102 & 58     & 31     & 0.23  & --0.15 & --0.42 & \checkmark\\
74 & 16:13:26   & 16:18:50 & 162 & 112    & 78     & --0.32 & /  & /  & \checkmark\\
75 & 16:16:26   & 16:21:50 & 92  & 46     & 39     & --0.25 & /  & /  & \checkmark\\
76 & 16:22:26   & 16:29:38 & 106 & 61     & 50     & 0.08  & --0.91 & /  & \checkmark\\
77 & 16:25:26   & 16:32:38 & 145 & 67     & 47     & --0.54 & --0.19 & /  &         \\
78 & 16:31:26   & 16:40:26 & 219 & 124    & 72     & --0.74 & --0.22 & /  & \checkmark\\
79 & 16:37:26   & 16:46:26 & 218 & 117    & 68     & --0.68 & --0.00 & /  & \checkmark\\
80 & 16:39:50   & 16:50:02 & 115 & 91     & 56     & --0.27 & --0.11 & /  & \checkmark\\
81 & 16:44:38   & 16:50:02 & 144 & 101    & 74     & --0.36 & /  & /  &         \\
\noalign{\smallskip}\hline
\end{longtable}
\twocolumn

\end{document}